# Direct visualization of 2D topological insulator in single-layer 1*T*'-WTe$_2$

Zhen-Yu Jia[1†], Ye-Heng Song[1†], Xiang-Bing Li[1], Kejing Ran[1], Pengchao Lu[1], Hui-Jun Zheng[1], Xin-Yang Zhu[1], Zhi-Qiang Shi[1], Jian Sun[1,2], Jinsheng Wen[1,2], Dingyu Xing[1,2], Shao-Chun Li[1,2]*

1. National Laboratory of Solid State Microstructures, School of Physics, Nanjing University, Nanjing, Jiangsu 210093, China.

2. Collaborative Innovation Center of Advanced Microstructures, Nanjing University, Nanjing 210093, China.

† These authors contributed equally to this work.

*e-mail: scli@nju.edu.cn

We have grown nearly freestanding single-layer 1T'-WTe$_2$ on graphitized 6*H*-SiC(0001) by using molecular beam epitaxy (MBE), and characterized its electronic structure with scanning tunneling microscopy / spectroscopy (STM/STS). The existence of topological edge states at the periphery of single-layer WTe$_2$ islands were confirmed. Surprisingly, a bulk band gap at the Fermi level and the insulating behaviors were also found in the single-layer WTe$_2$ at low temperature, which is likely associated with an incommensurate charge order transition. The realization of two-dimensional topological insulators (2D TIs) in single-layer transition metal dichalcogenide (TMD) provides a promising platform for further exploration of the 2D TIs' physics and related applications.

Two-dimensional topological insulators (2D TIs) feature the conducting edge states topologically protected from back scattering by the time reversal symmetry and the gapped bulk band structure, thus hosting the quantum spin Hall (QSH) effect [1-4]. This extraordinary property makes 2D TIs promising in future applications such as, low dissipation electronics, and quantum computing etc. Since the discovery of QSH effect in HgTe/CdTe quantum wells [5,6], increasing attentions have been gained. In particular, many theoretical predictions have been made, aiming for more practical QSH materials [7-9]. However, it still remains difficult to realize the QSH effect in a natural single-layer 2D material in a well-controlled fashion [10,11]. Recently, Reis *et al.* has achieved the Bismuthene on a SiC substrate, and observed both the conductance edge channels and the large band gap [12]. Although mechanical exfoliation has become the widely accepted method achieving the single-layer QSH materials [13,14], the samples have to be carefully safeguarded from contamination or reactions. Studies on 2DTI quantum wells have to suffer the sophisticated growth techniques [6,15,16]. Characterizations of topological band structures have been mostly focused on the surface of their bulk counterparts because of the limited availability of single-layer 2DTIs [10,17-20], in which the interlayer coupling has to be considered.

Recently, Qian *et al.* [9] predicted a family of large band gap QSH materials in the 1T' phase of transition metal dichalcogenide (TMD), $TX_2$, where T is a transition metal atom (Mo, W), and X a chalcogen atom (S, Se, or Te). These QSH materials exhibit the feasibility in the construction of van der Waals (vdW) devices [9]. Additionally, the 3D bulk form of $TX_2$, in particular $WTe_2$ and $MoTe_2$, also exhibit exotic physics such as the type-II Weyl semimetal and the unsaturated magnetoresistance [21,22]. Since TMDs boast outstanding properties such as chemically inert surface and weak interlayer interactions, the single-layer 1T' phase TMD is expected to find great potentials once the topological edge states are experimentally confirmed. A recent transport study has shown convincing evidence for conducting edge channels in exfoliated $1T'$-$WTe_2$ [13]. However, the real-space imaging of these edge states revealing, e.g., its spatial width, its relation to the edge geometry, and its relation to other spectral features of the 1T'-$WTe_2$ surface has not been published so far.

In this study, we have directly grown the single-layer 1T'-WTe$_2$ on graphitized 6$H$-SiC(0001) substrate by using molecular beam epitaxy (MBE) technique, and *in situ* characterized the band structure with atomic-scale scanning tunneling microscopy / spectroscopy (STM / STS). Through these measures, any possible contamination from the air or during transport was avoided, and the intrinsic electronic structure of monolayer WTe$_2$ was made available. The growth of monolayer WTe$_2$ was monitored by using reflection high-energy electron diffraction (RHEED). It has been found that the topological edge states are located along all kinds of the WTe$_2$ step edges. The measured STS results are consistent with the calculated band structure based on the many-body effect [9]. However, in contrast to the semimetal behavior as predicted by DFT [9], a bulk band gap is observed at the Fermi energy at low temperature, thus leading to the insulating behavior of the single-layer WTe$_2$. Concomitantly, an incommensurate charge order is found to occur at low temperature, which is likely to be associated with the gap opening.

The single-layer WTe$_2$ films were grown in ultrahigh vacuum with the base pressure of $1\times10^{-10}$ mbar. The 6H-SiC(0001) substrate was initially outgassed at ~600 °C, and then flashed up to ~1350 ºC for a few cycles until the BLG termination was acquired. The quality of the BLG/SiC(0001) was checked by RHEED and STM. High purity W (99.999%) and Te (99.999%) were evaporated from an e-beam evaporator and a Knudsen cell. The temperature of the substrate was kept at ~250 ºC during the WTe$_2$ growth, with a flux ratio of ~1:20 (W:Te). *In situ* STM/STS measurements were carried out with a low temperature STM (Unisoku USM1600) at ~4.7 K unless otherwise specified. STM measurements were performed with a constant current mode. A lock-in amplifier was used for STS measurements, with the ac modulation of 5-10 mV at 963 Hz. Electrical resistance measurements were conducted in a Physical Property Measurement System (PPMS) from Quantum Design. For the *ex-situ* resistance measurements, we deposited the Te capping layers on top of the single-layer WTe$_2$ film.

Typically, the freestanding single-layer WTe$_2$ takes the 1T' crystal structure, as sketched in Figure 1(a). The W atomic layer is sandwiched between the top and the bottom Te atomic layers [23]. To achieve a quasi-freestanding monolayer 1T'-WTe$_2$, coupling to the substrate has to be efficiently reduced, which might otherwise destroy the topologically nontrivial edge states. The epitaxial bilayer graphene (BLG) formed on the 6H-SiC(0001) substrate was used, due to its chemically inert surface and the vdW interlayer bonding [24]. WTe$_2$ prefers to grow in the layer-by-layer mode on BLG/SiC(0001) via optimizing the growth parameters, see the surface morphologies with various WTe$_2$ coverages in Supplementary Fig. S1 [25]. Figure 1(b) depicts the RHEED pattern of the single-layer WTe$_2$/BLG/SiC(0001). Pertaining to the symmetry registry between 1T'-WTe$_2$ (orthogonal) and graphene (hexagonal), there mainly exists three equivalent orientations of single-layer WTe$_2$ domains at the BLG surface. Randomly oriented WTe$_2$ islands are also occasionally observed, implying the weak coupling between WTe$_2$ and BLG/SiC(0001).

Figure 1(c) shows the typical STM topographic image of WTe$_2$ grown on BLG/SiC(0001) at a coverage of ~ 0.7 monolayers (MLs). The surface is dominated by the single-layer WTe$_2$ islands, with the exception of some small second-layer WTe$_2$ islands. Atomic resolution STM image shows clearly the locations of surface Te atoms, as marked by the colored dots on the inset to Figure 1(c). The step height of the first single-layer WTe$_2$ is ~1.21 nm, much larger than the step height for the bulk of 0.701 nm [23], thus indicating a large vdW gap between the BLG substrate and WTe$_2$ monolayer, see the line cut profile in Figure 1(d).

A typical differential conductance dI/dV curve measured on the terrace represents the bulk band structure of the single-layer WTe$_2$, as plotted at the bottom of Figure 1(e). For comparison, the dI/dV curves measured on top of the second-layer and the bulk WTe$_2$ are also plotted together. The STS for single-layer WTe$_2$ is visually dissimilar from the bulk, because the single-layer energy band is well separate and 2 dimensional, while the bulk energy band is continuous and 3 dimensional. The band edges for the first single layer, as tentatively assigned according to the DFT calculations [9], are in great agreement with the calculated band structure for freestanding single layer WTe$_2$ with many-body effect (see Fig. S1(F) in Ref [9]). Such a nice agreement, particularly in this large energy range between ±1V [9], further confirms the weakly coupling of WTe$_2$ to BLG. The notable exception is that the density of states is drastically suppressed at the Fermi energy, indicating a gap is opened, see Figure 1(f). In fact, the dI/dV intensity inside the gap is not suppressed to zero [see Figure 1(f)], indicating it is not a full gap. High resolution dI/dV spectrum [Figure 1(g)] shows the coherence peaks at the gap edges. Such a gap is found ranging from ~20 to ~40 mV on the surface. The gap is most prominent in the first single layer, less in the second layer, and disappears in the bulk. The origin of the gap will be discussed below. It is worthwhile noting that the intensity between ~-0.5 eV to ~+0.5 eV is rather low for the first single-layer $WTe_2$, even though there exist non-zero density of states.

Spatially resolved differential conductance dI/dV spectrum represent the variation of local density of states at the $WTe_2$ surface. The electronic structure near the periphery of the $WTe_2$ islands is disclosed in a series of dI/dV spectra taken along the line spanning from the step edge to the terrace, as shown in Figure 2(a). The dI/dV intensity gradually increases as it approaches to the step edge, especially in the bias range between ~-0.5 eV to ~+0.5 eV. dI/dV spectra are also taken in a smaller bias energy range, from –300 mV to +300 mV, see Figure 2(b). The feature located at ~ -40 mV is found to only exist near the step edge in a spatial range of ~5.0 nm, as marked by the red vertical line in Fig. 2(b).

Various orientations of the step edges are found in the periphery of the $WTe_2$ monolayer islands, which is different from the previous study for another 2D TI $ZrTe_5$ [20]. In fact, the enhancement of the dI/dV intensities taking place along the step edges subtly vary, suggesting that the edge states are robust to the weak perturbations at the steps. Figures 2(c) and (d)-(h) show the topographic image and the corresponding dI/dV maps taken at various bias energies, respectively. Evidently, the edge states, as indicated by the enhanced dI/dV intensity, are always distributed along the $WTe_2$ steps. The penetration length of the edge state is estimated to be ~5.0 nm, as shown in Figure 2(i), in line with the predicted value given by Qian *et al* [9]. Occasionally, the dI/dV intensities around the step edges exhibit obvious discontinuity at certain bias energies, particularly in the upper left corner of Fig 2(f) and (g). Such discontinuity in the dI/dV maps may be caused by the hybridization between the edge states with the bulk states out of the bulk band gap.

Next, we continue to discuss the origin of the bulk band gap in the single-layer $WTe_2$. To investigate the transport behavior when such a band gap is opened, the temperature-dependent resistance is measured over a full layer of the single-layer WTe$_2$ and plotted in Figure 3(a), see supplementary Fig. S1(e,f) [25] for the corresponding surface morphologies. The resistance exhibits a sudden increase followed by a kink as the temperature decreases to below ~100 K, and another insulating kink at ~30 K. For comparison, the resistance of the pure BLG/SiC(0001) substrate is also measured, as shown in Figure 3(a). The whole behavior of WTe$_2$/BLG/SiC(0001) from 300 K down to ~50 K might be easily attributed to the BLG/SiC(0001), while only the stronger insulating kink below ~30 K must be related to the 1$T$'-WTe$_2$. To exclude the possible contribution from the BLG/SiC(0001) substrate, a two-channel model is assumed, as shown in the inset to Figure 3(a). The measured conductance, $G_{Total} = G_{WTe_2} + xG_{BLG}$, and $0 \leq x < x_m$ presents the BLG/SiC substrate effect, with $x = 0$ standing for no substrate effect and $x = x_m$ the maximal substrate effect. It then follows that the effective resistance of WTe$_2$, $R_{WTe_2} = R_{Total}/(1 - x R_{Total} / R_{BLG})$. Figure 3(b) shows the extracted $R_{WTe_2}$ with various $x$ values. Regardless of the values of $x$ factor, it is evident that the WTe$_2$ single layer behaves as an insulator at low temperature (T < ~30 K).

Figure 4 shows the typical STM topographic images and the corresponding fast Fourier transforms (FFT) taken at negative biases at various temperatures. In addition to the Bragg peaks corresponding to the lattice unit along x and y axis, as marked by red circles, extra peaks are clearly identified at ~4.7 K, as marked by the blue circles. Such a FFT peak corresponds to a spatial charge periodicity of ~3.3 times of the unit size along y axis. Such a charge order can persist up to LN$_2$ temperature, but disappears at room temperature, indicating that a charge order transition may occur between LN$_2$ and room temperature, as characterized in both STM images and the FFT. In general, QPI occurring at the step edge induces the spatial charge modulations, but with a periodicity that is dependent on the bias energy. As seen in the topographic image and the corresponding FFT in Fig. 4, contributions from the tunneling electrons are integrated in the range from Fermi level to the bias voltage, thus the QPI-induced charge modulation appear smeared. Furthermore, the FFT transforms of the topographic images at different bias voltages indicate that the corresponding wave vector in the reciprocal space does not change as the bias energy varies. Therefore, it is concluded that there exists a static charge order.

All the experimental results point to the conclusion that the single-layer 1*T*'-WTe$_2$ undergoes an incommensurate charge order transition and thus opens a band gap at the Fermi energy, turning the topological semimetal into topological insulator at low temperature. The transition temperature for the incommensurate charge order obtained from STM data seems discrepant with that obtained from the transport result, which might be related to the multiple CDW transitions as in TaS$_2$ or to the Te capping. Furthermore, the band gap and the coherence peaks can be well fitted with a CDW model [26], see Figure 1(g). The coincidence of the periodicity of the incommensurate charge order with the electron pockets located at ±Λ [9] implies that the electron-phonon or electron-hole interactions may play a role in driving such a transition, which are expected to be greatly enhanced in the single-layer limit of 2D materials [27]. Further in-depth investigation is required to uncover the mechanism for the charge order transition. In addition, recent theoretical studies also predicted a band gap as well when considering the many-body effect [14] or the uniaxial strain [28].

In summary, we directly grow the single-layer 1*T*'-WTe$_2$ and *in situ* visualize its topological edge states. The single-layer WTe$_2$ is weakly coupled to the BLG/SiC(0001) substrate, thus exhibiting a nearly freestanding electronic structure. The incommensurate charge order transition at low temperature opens a bulk band gap at the Fermi energy, thus making the single-layer WTe$_2$ a true 2D TI. In addition, the local density of state is extremely low for the bulk band structure in a larger energy scale, particularly between ~-0.5 eV - ~+0.5 eV, and thus it is expected that the fermions of the topological edge states dominate over the band fermions in such a large energy scale, which may lead to a prominent QSH effect at low temperatures. The epitaxial monolayer WTe$_2$ grown by MBE provides a contamination-free material base for the QSH study and also facilitate the device construction by vdW epitaxy of 2D functional materials.

**ACKNOWLEDGMENTS**

This work was supported by the Ministry of Science and Technology of China (Grants No.

2014CB921103, No. 2013CB922103, No. 2016YFA0300404, No. 2015CB921202), and by the National Natural Science Foundation of China (Grants No. 11374140, No. 11374143, No. 11674157, No. 51372112, No. 11574133), and NSF Jiangsu Province (No. BK20150012).

Z.-Y. J. and Y.-H. S. contributed equally to this work.

Note added: During the referee process we became aware two other relevant manuscripts [29, 30]

**FIGURE CAPTIONS**

FIG. 1 Single-layer WTe$_2$ islands grown on BLG/SiC(0001). (a) Top and side view sketches of single-layer 1$T$'-WTe$_2$ on BLG/SiC(0001); (b) RHEED pattern of the as grown single-layer WTe$_2$. The blue arrow marks the streaks from the BLG/SiC(0001) substrate, while the red arrows represent the ones from WTe$_2$ domains of three equivalent orientations. (c) Large-scale STM topographic image of single-layer WTe$_2$ islands on BLG/SiC(0001) substrate (150 × 150 nm$^2$, $U$ = +1.0 V, $I_t$ = 100 pA, T = ~4.7 K). The length of the scale bar is 30 nm. Inset: Atomic-resolution STM image of the single-layer WTe$_2$ surface ($U$ = +50 mV, $I_t$ = 500 pA, T = ~4.7 K) showing that the single-layer WTe$_2$ is in 1$T$' phase. The red rectangle represents the surface unit cell. (d) Line scan profile taken across both the first and the second layer WTe$_2$. (e) dI/dV spectra taken on surfaces of the single-layer WTe$_2$ (red), the second-layer WTe$_2$ (green) and the bulk termination (blue), respectively. The labels C2, C3 and V2, V3 mark the edges of valence bands and conduction bands, according to the calculated band structure [9]. (f) Small energy scale (from -300 mV to +300 mV) dI/dV spectra. (g) High-resolution dI/dV spectrum (red) taken at ~6.7 K indicating the features of the bulk band gap. Both the band gap and the coherence peaks are fitted (black dashed) based on a CDW gap format adopted from the literature [26].

FIG. 2 Edge states of the single-layer 1$T$'-WTe$_2$ on BLG/SiC(0001). (a) Large energy scale dI/dV spectra (from -1.0V to +1.0V) taken along the rainbow-colored dotted line in the inset at ~4.7 K, sweeping from the terrace (red end) to the step edge (purple end). The inset shows the STM topographic image ($U$ = +500 mV, $I_t$ = 200 pA) where the dI/dV spectra were taken. The colored dots mark the locations where the same colored spectra were taken, respectively. (b) dI/dV spectra (from -300 mV to +300 mV) taken across the step edge. The corresponding line scan profile of the step edge is plotted in the inset, which is parallel to the *y* axis. The edge-state feature is marked by the red vertical line, and the bulk band gap the black vertical lines. (c) STM topographic image (28 × 28 nm$^2$; $U$ = +1.0 V; $I_t$ = 100 pA). The length of the scale bar is 5.6 nm. (d)-(h) dI/dV maps measured on the same area of (c) with various bias voltages of +200 mV, +100 mV, +60 mV, -40 mV and -80 mV respectively. (i) The typical line cuts extracted from (d)-(g) indicating the penetration length of the edge state is ~5.0 nm.

FIG. 3 Transport behavior of single-layer 1T'-WTe$_2$. (a) The measured square resistance $R_{sq}$ as a function of temperature. The black dotted line is $R_{Total}$ as measured on the Te capped WTe$_2$/BLG/SiC, and the red dotted line $R_{BLG}$ as measured on the Te capped BLG/SiC. The inset shows the schematic illustration of the resistance measurement. (b) The extracted effective resistance of single-layer WTe$_2$, $R_{WTe2}$, assuming $G_{Total} = xG_{BLG} + G_{WTe2}$, with various values of $x$. Here $0 \leq x < x_m$ indicates the BLG/SiC substrate effect on $G_{Total}$, with $x=0$ standing for no substrate effect and $x= x_m$ corresponds to the maximal substrate effect.

FIG. 4 Incommensurate charge order in the single-layer 1T'-WTe$_2$ at low temperature. (a)-(c) STM topographic images (left panel) and the corresponding FFT (right panel) taken at ~4.7K, LN$_2$ and RT, respectively. The bias voltages $U$ are marked in the STM images, and $I_t$ = 0.1 nA. The red circles in the FFT mark the Bragg vectors, and the blue ones the wave vector for the incommensurate charge order.

[19] C. Pauly, B. Rasche, K. Koepernik, M. Liebmann, M. Pratzer, M. Richter, J. Kellner, M. Eschbach, B. Kaufmann, L. Plucinski, C. M. Schneider, M. Ruck, J. van den Brink, and M. Morgenstern, Nature Phys. **11**, 338 (2015).

[20] X. B. Li, W. K. Huang, Y. Y. Lv, K. W. Zhang, C. L. Yang, B. B. Zhang, Y. B. Chen, S. H. Yao, J. Zhou, M. H. Lu, L. Sheng, S. C. Li, J. F. Jia, Q. K. Xue, Y. F. Chen, and D. Y. Xing, Phys. Rev. Lett. **116**, 176803 (2016).

[21] M. N. Ali, J. Xiong, S. Flynn, J. Tao, Q. D. Gibson, L. M. Schoop, T. Liang, N. Haldolaarachchige, M. Hirschberger, N. P. Ong, and R. J. Cava, Nature **514**, 205 (2014).

[22] A. A. Soluyanov, D. Gresch, Z. Wang, Q. Wu, M. Troyer, X. Dai, and B. A. Bernevig, Nature **527**, 495 (2015).

[23] P. Lu, J.-S. Kim, J. Yang, H. Gao, J. Wu, D. Shao, B. Li, D. Zhou, J. Sun, D. Akinwande, D. Xing, and J.-F. Lin, Phys. Rev. B **94**, 224512 (2016).

[24] C. Berger, Z. Song, X. Li, X. Wu, N. Brown, C. Naud, D. Mayou, T. Li, J. Hass, A. N. Marchenkov, E. H. Conrad, P. N. First, and W. A. de Heer, Science **312**, 1191 (2006).

[25] Supplementary Materials.

[26] U. Chatterjee, J. Zhao, M. Iavarone, R. Di Capua, J. P. Castellan, G. Karapetrov, C. D. Malliakas, M. G. Kanatzidis, H. Claus, J. P. Ruff, F. Weber, J. van Wezel, J. C. Campuzano, R. Osborn, M. Randeria, N. Trivedi, M. R. Norman, and S. Rosenkranz, Nature Commun. **6**, 6313 (2015).

[27] X. Xi, L. Zhao, Z. Wang, H. Berger, L. Forro, J. Shan, and K. F. Mak, Nature Nanotech. **10**, 765 (2015).

[28] H. Xiang, B. Xu, J. Liu, Y. Xia, H. Lu, J. Yin, and Z. Liu, AIP Advances **6**, 095005 (2016).

[29] S. Tang, C. Zhang, D. Wong, Z. Pedramrazi, H.-Z. Tsai, C. Jia, B. Moritz, M. Claassen, H. Ryu, S. Kahn, J. Jiang, H. Yan, M. Hashimoto, D. Lu, R. G. Moore, C. Hwang, C. Hwang, Z. Hussain, Y. Chen, M. M. Ugeda, Z. Liu, X. Xie, T. P. Devereaux, M. F. Crommie, S.-K. Mo, Z.-X. Shen, arXiv:1703.03151.

[30] L. Peng, Y. Yuan, G. Li, X.g Yang, J.-J. Xian, C.-J. Yi, Y.-G. Shi, Y.-S. Fu, arXiv:1703.05658.
12

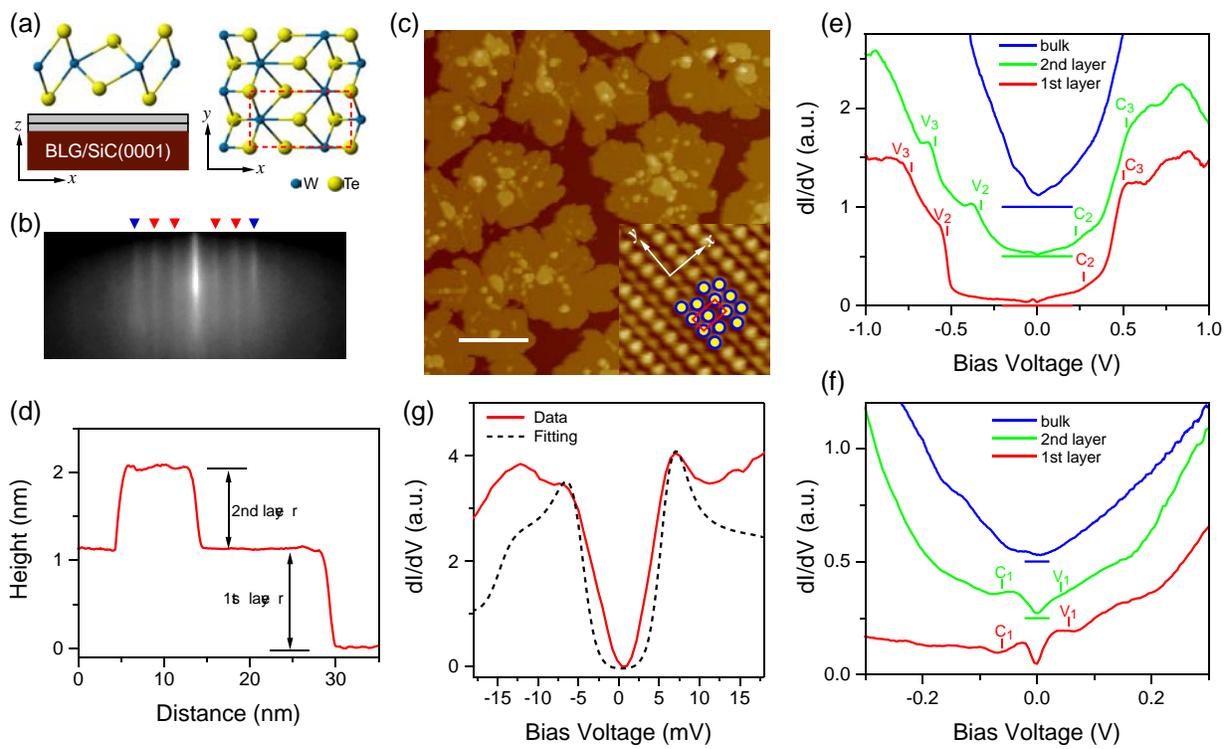

Figure 1

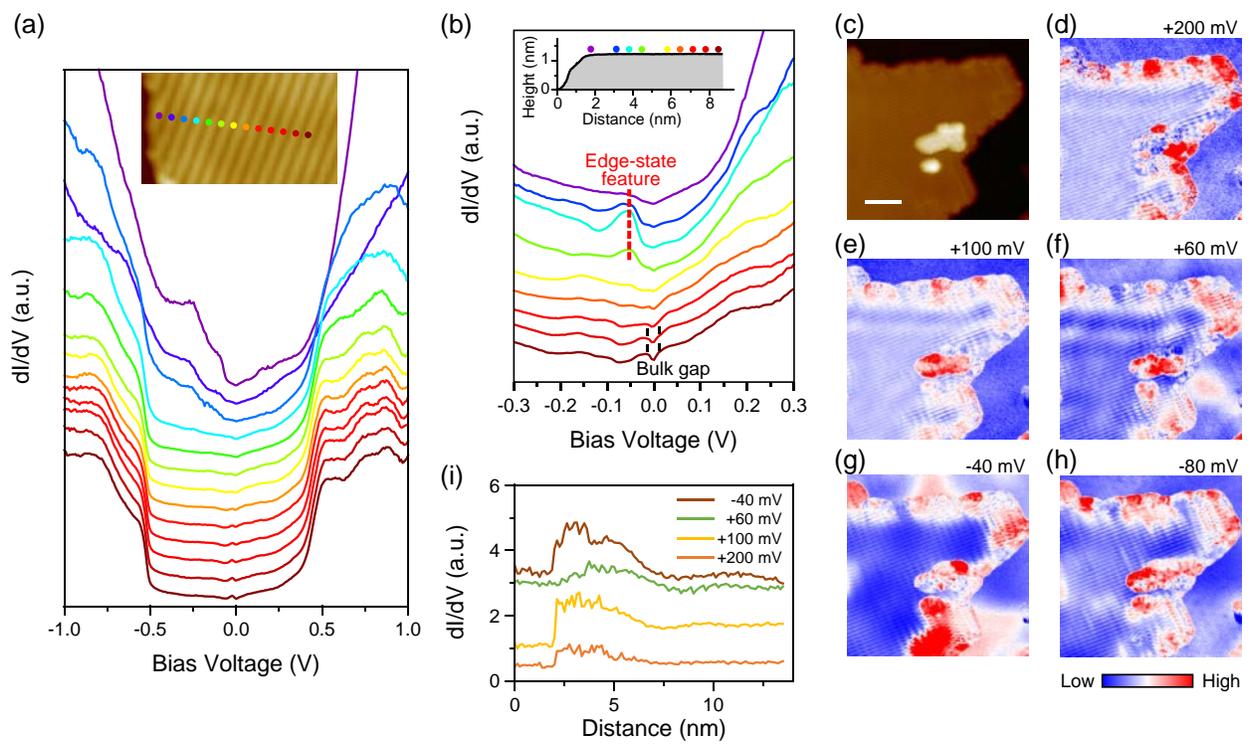

Figure 2

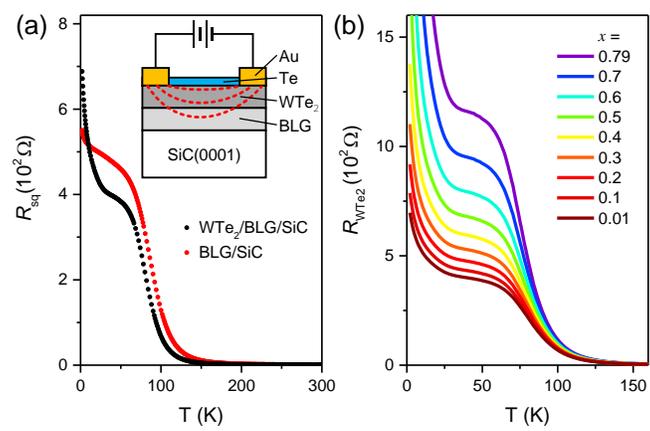

Figure 3

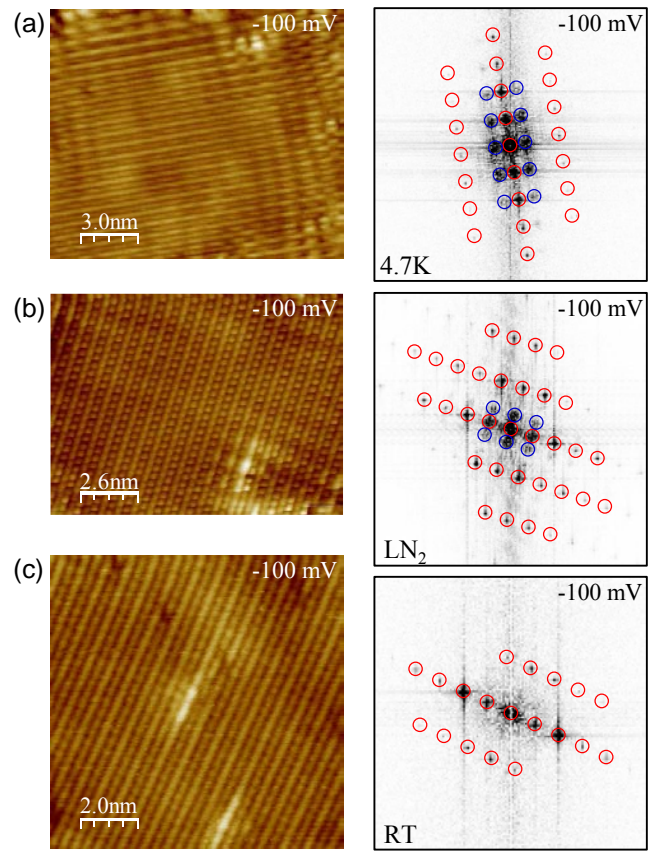

Figure 4